\documentclass[10pt]{article}

\begin{document}

\title{Modern cosmologies from  empty Kaluza-Klein  solutions in 5D}
\author{J. Ponce de Leon\thanks{E-Mail:
jpdel@ltp.upr.clu.edu, jpdel1@hotmail.com}  \\Laboratory of Theoretical Physics, 
Department of Physics\\ 
University of Puerto Rico, P.O. Box 23343,  
San Juan,\\ PR 00931, USA}
\date{February 13,  2009}

\maketitle
\begin{abstract}
We show that the empty five-dimensional  solutions of Davidson-Sonnenschtein-Vozmediano,  {\em Phys. Rev.} {\bf D32} (1985)1330, in the ``old" Kaluza-Klein gravity, under appropriate interpretation can generate an ample variety of cosmological models in 4D, which include the higher-dimensional modifications to general relativity  predicted by ``modern" versions of noncompactified 5D gravity as, e.g., induced-matter and braneworld theories. This is the first time  that these solutions are investigated in a systematic way as embeddings for  cosmological models in 4D. They provide a different formulation,  which   is complementary to the approaches used in current versions of 5D relativity.

\end{abstract}

\bigskip



\newpage
\section{Introduction}

The concept that  our universe might be a 4-dimensional hypersurface embedded in a higher-dimensional universe constitutes an extremely provocative idea, from a theoretical, philosophical  and practical point of view.

From a theoretical point of view, the well-known Campbell's theorem \cite{Campbell}, \cite{Seahra} serves as a ladder to go between manifolds whose dimensionality differs by one. This theorem, which is valid in any number of dimensions, implies that every solution of the 4D Einstein equations with arbitrary energy-momentum tensor can be embedded, at least locally, in a solution of  the 5D  Einstein field equations in vacuum. It is the backbone of induced-matter  theory (IM) \cite{general}, and brings to fruition the philosophy of geometrodynamics \cite{Feynman} in which ``matter and charge may be manifestations of the topology of space\footnote{As  Feynman put it: ``It would indeed be very beautiful to have $G^{\mu}_{\nu}$ = 0 everywhere, so that, in words used recently to describe geometrodynamics, matter comes from no matter, and charge comes from no charge" \cite{Feynman}.} 

From a practical point of view, extensions of general relativity to five and more dimensions seem to provide the best route to unification of gravity with interactions of particle physics \cite{Davidson Owen}-\cite{particle physics}. Braneworld theory (BW) proposes a model where our spacetime is a singular hypersurface embedded in an empty (no matter sources) 5-dimensional anti-de Sitter space \cite{algebraically}, which  might provide a possible solution to the hierarchy problem between weak and Plank scales. 

One important physical problem in higher-dimensional theories is to develop a full understanding of implications in 4D. Therefore, it is essential to compare and contrast the effective  pictures generated in 4D by different  versions  of   5-dimensional relativity, like, e.g.,  the classical Kaluza-Klein theory with ``cylinder condition" (KK) and the two above-mentioned approaches, where the extra dimension is not assumed to be compactified.

Today, it is well known that  IM and BW originate the same effective 4-dimensional world, despite of the fact  that they have different motivation and interpretation \cite{physical motivation}. For example, in cosmological applications,  on {\it every} 4D hypersurface orthogonal to the extra dimension IM reproduces, although in different conventions and notation,  the   generalized (or modified) Friedmann equation of BW on a ${{\bf{Z}}}_{2}-$symmetric brane. 
 
The question arises of whether one can establish some connection between cosmological models constructed under similar conditions in KK, IM, and BW. 
The aim of this work  is  to study this question. In particular, we ask whether models in the ``old" Kaluza-Klein theory can be made compatible with IM and BW, i.e., with the concept that our universe is a  4D hypersurface embedded in a 5-dimensional world. 

We will see that the answer to this question is positive.
In our discussion we concentrate our attention to a family of KK cosmological models 
first discovered by Davidson, Sonnenschtein and Vozmediano \cite{DSV}. These models  share with IM and BW the property  that they are solutions to the Einstein field equations in an empty 5-dimensional space 
\begin{equation}
\label{5D equations with Lambda5}
{^{(5)}G}_{AB} = k_{(5)}^2\Lambda_{(5)}\gamma_{AB},
\end{equation}
where $A, B = 0, 1, 2, 3, 4$; $k_{(5)}^2$ is a constant introduced for dimensional considerations, and  $\gamma_{AB}$ is the extended $5D$ Friedmann-Robertson-Walker (FRW) metric 
\begin{equation}
\label{five-dimensional RW metric}
d{\cal{S}}^2 = \gamma_{AB} dx^A dx^B = dt^2 - A^2(t)d\Sigma_{k}^2 - \Phi^2(t)dy^2. 
\end{equation}
Here $d\Sigma_{k}^2$ is the metric on the unit three dimensional plane $(k = 0)$, hyperboloid $(k = - 1)$ or sphere $k = 1$, viz.,
\begin{equation}
d \Sigma_{k}^2 = \frac{dr^2}{1 - k r^2} + r^2\left(d\theta^2 + \sin^2\theta d\phi^2\right).
\end{equation}

In our notation,  the Davidson-Sonnenschtein-Vozmediano (DSV) solutions  are given by\footnote{Some typos in  the original paper \cite{DSV} are fixed here.}

\begin{equation}
\label{General Davidson el al solution}
A^2(t) = \left\{\begin{array}{cc}
              c_{1} \cosh{\omega t} + c_{2} \sinh{\omega t}  + \frac{2 k}{\omega^2}, \;\;\;\mbox{for}\;\;\;k_{(5)}^2\Lambda_{(5)} = \frac{3\omega^2}{2},\\
\\
- k t^2 + c_{1} t + c_{2},\;\;\;   \mbox{for}\;\;\; k_{(5)}^2\Lambda_{(5)} = 0,  \\
\\
               c_{1}\cos{\omega t} + c_{2} \sin{\omega t} - \frac{2 k}{\omega^2},\;\;\;    \mbox{for} \;\;\; k_{(5)}^2\Lambda_{(5)} = - \frac{3 \omega^2 }{2},
               \end{array}
              \right.
\end{equation}
\begin{equation}
\label{Phi for the general Davidson solution}
\Phi(t) = B_{0}\left(\frac{d A}{dt}\right),
\end{equation}
where $c_{1}$, $c_{2}$,  $B_{0}$ are constants of integration with the appropriate dimensions\footnote{$B_{0}$ has dimensions of length. In the first and third solutions  $c_{1}$ and $c_{2}$ are dimensionless. In the second solution $c_{1}$ has dimensions of (length)$^{- 1}$ and $c_{2}$ is dimensionless}.   To study the effective 4-dimensional world, these authors used the Kaluza-Klein ansatz, which consists in assuming that  the extra dimension is  compactified (rolled up to a small size), and identifying 

\begin{equation}
\label{4D metric in Kaluza-Klein with cilindricity}
 g_{\mu\nu} = \Phi \gamma_{\mu\nu}, 
\end{equation}
with the metric of the effective 4-dimensional world \cite{Dolan}.   An observer in 4D, who is not aware of the existence of an extra dimension, interprets the metric $g_{\mu\nu}$ as if it were governed by an effective 4D energy-momentum tensor.  For the  DSV solutions, the effective  isotropic pressure $p^{\mbox{eff}}$ and density $\rho^{\mbox{eff}}$ satisfy the ``equation of state"
\begin{equation}
\label{equation of state for Davidson solution}
\rho^{\mbox{eff}} = p^{\mbox{eff}} + \frac{k_{(5)}^2\Lambda_{(5)}}{4\pi G \Phi}, 
\end{equation}
 which for $\Lambda_{(5)} = 0$ reduces to the familiar stiff equation.

However, this is not the only way to establish the effective 4-dimensional picture. Other alternatives  for dimensional reduction  are formulated in  induced-matter and braneworld theories, where the extra dimension is not assumed to be compactified.  Since the DSV solutions are vacuum solutions, it is of theoretical interest to reanalyze their 4-dimensional interpretation from the perspective  of these theories. 

In section 2, we sketch the main features of noncompactified theories. We show that, by means of a simple transformation $t\leftrightarrow y$,  from  (\ref{General Davidson el al solution})-(\ref{Phi for the general Davidson solution}) one can  
generate a family of static 5-dimensional solutions. Then, within the context of IM in the comoving frame,  we demonstrate that the DSV solutions and their static counterparts lead to significantly different scenarios in 4D.
In section $3$, we study the 4-dimensional picture measured by an observer who, instead of being at rest, is moving in a DSV universe. We find  that such an observer can perceive a rich variety of cosmological scenarios, including cosmological models where the induced matter satisfies the barotropic equation of state. 
In section $4$,  we show that the static DSV models  allow us to reproduce the modified, or generalized, Friedmann cosmological equation of branewold models \cite{Binetruy}-\cite{Ida}.

\section{Relaxing the Kaluza-Klein ansatz}

 Modern theories of gravity in 5D introduce two important new ingredients. First, they   do not require the  extra dimension to be compact; in principle it can be infinitely large.  Second, a large extra dimension can be either spacelike or timelike; both are physically admissible (see, e.g., \cite{signature} and references therein). In this regard, one should be careful to discriminate between temporal (spatial) dimensions, which actually have physical units of time (length); and timelike (spacelike) ones, which merely have timelike (spacelike) signature \cite{Overduin Wesson}. 

\subsection{Static solutions}

An immediate consequence of these new ``ingredients"  is that  the metric 

\begin{equation}
d{\cal{S}}^2 = d t^2 - A^2(t) d\Sigma_{k}^2 + \epsilon B_{0}^2\left(\frac{d A}{dt}\right)^2 dy^2, 
\end{equation}
with timelike extra dimension $(\epsilon = + 1)$, is also a physically valid solution of the field equations $G_{AB} = k_{(5)}^2\Lambda_{(5)}\gamma_{AB}$. In this case making the transformation $t \leftrightarrow y$  and $k \rightarrow \epsilon k$,  from (\ref{General Davidson el al solution})-(\ref{Phi for the general Davidson solution}) we obtain the following set of {\it static} solutions 

\begin{equation}
\label{static solution}
d{\cal{S}}^2 = \gamma_{CD}dx^C dx^D = B_{0}^2 \left(\frac{d{\cal{A}}}{dy}\right)^2 dt^2 - {\cal{A}}^2 d\Sigma_{k}^2 + \epsilon dy^2,
\end{equation}
where
\begin{equation}
\label{Transformed General Davidson el al solution}
{\cal{A}}^2(y) = \left\{\begin{array}{cc}
              c_{1} \cosh{\omega y} + c_{2} \sinh{\omega y}  + \frac{2 \epsilon k}{\omega^2},    \;\;\;k_{(5)}^2\Lambda_{(5)} = \frac{3\epsilon \omega^2}{2}\\
\\
  - \epsilon k y^2 + c_{1} y + c_{2},\;\;\;   \mbox{for}\;\;\;k_{(5)}^2\Lambda_{(5)} = 0, \\
\\
   c_{1}\cos{\omega y} + c_{2}\sin{\omega y} - \frac{2 \epsilon k}{\omega^2}\;\;\; \mbox{for} \;\;\; k_{(5)}^2\Lambda_{(5)} = - \frac{3 \epsilon \omega^2 }{2}.
               \end{array}
 \right.
\end{equation}
We will show in sections 3 and 4 that these solutions, which we will call static DSV solutions,  may be used to establish a connection between the DSV solutions, the FRW models of conventional 4D relativity, and the modified Friedmann equation.
Although we are not especially promoting  timelike extra dimensions, because they can lead to closed timelike curves (CTC) and hence allow causality violation\footnote{It has been argued that physics can be compatible with CTC \cite{Friedman}, \cite{Bonnor}. For theories with timelike extra dimensions see, e.g., \cite{Twotimes}-\cite{{Gogber2}}.}, for the sake of generality in our discussion we keep $\epsilon = \pm 1$.

\subsection{Effective gravity in 4D}
 
The effective metric measured by an observer depends on her/his state of motion. The simplest physical scenario emerges in the rest  (also called {\it comoving}) frame, which in the present case means $(d x^i = dy = 0)$. 
In such a frame, the spacetime is recovered by going onto some hypersurface $\Sigma_{y_{0}}:y = y_{0} =$ constant, which is  orthogonal to the unit 5D vector $\hat{n}_{A} = \Phi \delta_{A}^{4}$ tangent to the extra coordinate.  The effective equations for gravity in $\Sigma_{y_{0}}$ are obtained from dimensional reduction of the 5-dimensional Einstein field equations, which is based on Campbell's theorem. It   consists in isolating the 4D part of the relevant 5D geometric quantities and use them to construct the 4D Einstein tensor ${^{(4)}G}_{\alpha \beta}$. For the 5D metric 
$d{\cal{S}}^2 = \gamma_{\mu\nu}dx^{\mu}dx^{\nu}  + \epsilon \Phi^2 dy^2$ the result is\footnote{There are five more equations; in the present case they reduce to $E_{\mu}^{\mu} = 0$ and ${^{(4)}T}^{\mu}_{\nu; \mu} = 0$.} 

\begin{eqnarray}
\label{4D Einstein with T and K}
{^{(4)}G}_{\alpha\beta} = 8 \pi G \;{^{(4)}T}_{\alpha\beta} \equiv &&\frac{2}{3}k_{(5)}^2\left[{^{(5)}{T}}_{\alpha \beta} + ({^{(5)}{T}}_{4}^{4} - \frac{1}{4}{^{(5)}{T}})g_{\alpha \beta}\right] - \nonumber \\
 &&\epsilon\left(K_{\alpha\lambda}K^{\lambda}_{\beta} - K_{\lambda}^{\lambda}K_{\alpha\beta}\right) + \frac{\epsilon}{2} g_{\alpha\beta}\left(K_{\lambda\rho}K^{\lambda\rho} - (K^{\lambda}_{\lambda})^2 \right) - \epsilon E_{\alpha\beta}, 
\end{eqnarray}
where ${^{(4)}G}_{\alpha\beta}$ is calculated with the 4D metric\footnote{Various versions of IM can be found in the literature for different definitions of the physical metric in 4D (see, e.g., \cite{various IM} and references therein).} $g_{\mu\nu} = \gamma_{\mu\nu}$; ${^{(4)}T}_{\alpha\beta}$ is the effective energy-momentum tensor (EMT) measured in $\Sigma_{y_{0}}$; ${^{(5)}{T}}_{AB}$ is the EMT in 5D;        $\epsilon = \pm 1$ depending on whether the extra dimension is spacelike or timelike;  $K_{\mu\nu}$ is the extrinsic curvature of $\Sigma_{y_{0}}$,
\begin{equation}
\label{extrinsic curvature}
K_{\alpha\beta} = \frac{1}{2}{\cal{L}}_{\hat{n}}g_{\alpha\beta} = \frac{1}{2\Phi}\frac{\partial{g_{\alpha\beta}}}{\partial y};  
\end{equation}
$E_{\mu\nu}$ is the projection of the 5D Weyl tensor ${^{(5)}C}_{ABCD}$ orthogonal to ${\hat{n}}^A$, i.e., ``parallel" to $\Sigma_{y_{0}}$, viz.,
\begin{equation}
\label{Weyl Tensor}
E_{\alpha\beta} = {^{(5)}C}_{\alpha A \beta B}{\hat{n}}^A{\hat{n}}^B 
= - \frac{1}{\Phi}\frac{\partial K_{\alpha\beta}}{\partial y} + K_{\alpha\rho}K^{\rho}_{\beta} - \epsilon \frac{\Phi_{\alpha;\beta}}{\Phi},
\end{equation}
and $\Phi_{\alpha} \equiv \partial \Phi/\partial x^{\alpha}$. In what follows we denote ${^{(4)}}T_{0}^{0} \equiv \rho^{\mbox{eff}}$ and ${^{(4)}}T_{1}^{1} = {^{(4)}}T_{2}^{2}=  {^{(4)}}T_{3}^{3} \equiv - p^{\mbox{eff}}$. 

\subsection{Interpretation of DSV solutions on $\Sigma_{y_{0}}$}

We now apply the above expressions to the original DSV solutions. Since the metric in (\ref{General Davidson el al solution})-(\ref{Phi for the general Davidson solution}) is independent of $y$, the extrinsic curvature $K_{\alpha \beta}$ of hypersurfaces $y = y_{0}$ is identically zero. Considering that  $E_{\mu\nu}$ is traceless,  the effective matter in $\Sigma_{y_{0}}$  can be interpreted as  a mixture of vacuum fluid and (Weyl) radiation. For future purposes we give the explicit form of the induced matter quantities in $\Sigma_{y_{0}}$: 
\begin{equation}
\label{vacuum fluid plus radiation}
\rho^{\mbox{eff}} = \rho + \frac{\Lambda_{(4)}}{8\pi G}, \;\;\;p^{\mbox{eff}}= p - \frac{\Lambda_{(4)}}{8 \pi G}, \;\;\;\Lambda_{(4)} = \frac{3 \epsilon_{\Lambda}\omega^2}{4}, \;\;\;p = \frac{\rho}{3},
\end{equation}
where $\epsilon_{\Lambda} = (1, 0, - 1)$ for $\Lambda_{(5)} > 0$, $\Lambda_{(5)} = 0$ and $\Lambda_{(5)} < 0$, respectively, and 
\begin{equation}
\label{The function fsubLambda}
8\pi G \rho = \frac{3 f_{\Lambda}}{4 A^4(t)}, \;\;\;\;f_{\Lambda} = \left\{\begin{array}{cc}
              \omega^{- 2}\left[4k^2 - \omega^4(c_{1}^2 - c_{2}^2)\right],    \;\;\;\Lambda_{(5)} > 0,\\
\\
  c_{1}^2 + 4k c_{2},\;\;\;   \Lambda_{(5)} = 0, \\
\\
   \omega^{- 2}\left[\omega^4(c_{1}^2 + c_{2}^2) - 4 k^2\right],\;\;\; \Lambda_{(5)} < 0.
               \end{array}
 \right.
\end{equation}
In cosmological applications, this interpretation  seems to be more satisfactory than the one given by the equation of state (\ref{equation of state for Davidson solution}),  derived from the Kaluza-Klein ansatz. We note that, in general,  $\rho$ is not necessarily positive, except for $k = 0$ and $\Lambda_{(5)} \leq 0$. However, if we add the initial condition $A(0) = 0$ then $\rho \geq 0$  in all three cases, regardless of the choice of $k$.

Let us now turn our attention to the static solutions (\ref{static solution})-(\ref{Transformed General Davidson el al solution}).  On every hypersurface $\Sigma_{y_{0}}$ the metric functions in (\ref{static solution}) are constants,  and the line element is Minkowskian. However, the components of the extrinsic curvature (\ref{extrinsic curvature}) are not zero, and the matter variables do not vanish in general. In fact,   
in all three cases,  the effective matter induced in $\Sigma_{y_{0}}$ is given by
\begin{equation}
\label{embedding with Y = constant}
p^{\mbox{eff}} = - \frac{\rho^{\mbox{eff}}}{3}, \;\;\;\; 8\pi G \rho^{\mbox{eff}} = \frac{3 k}{{\cal{A}}^2(y_{0})}.
\end{equation}
Thus, the relationship between the effective quantities is similar to the equation of state $\rho = - 3p$ for ``nongravitating matter", which has been discussed in a number of different contexts \cite{Davidson Owen}, \cite{Gott}-\cite{WessonEssay}. The nonvanishing components of the Riemann tensor on this hypersurface are 
\begin{equation}
\label{Riemann tensor for the static solution}
R_{1313} = R_{1212}\sin^2\theta = \frac{1 - k r^2}{r^2}R_{2323} = - \frac{k {\cal{A}}^2(y_{0}) r^2 \sin^2\theta}{1 - kr^2}.
\end{equation}
Thus, on $\Sigma_{y_{0}}$ the effective spacetime is empty and Riemann-flat only for $k = 0$. A similar, but not identical, {\it non-vacuum Minkowskian} spacetime is discussed in \cite{Campos Maartens}.

\section{Generating spacetime on a dynamical hypersurface}

Thus, when we identify our spacetime with some hypersurface $y = y_{0} =$ constant in a DSV universe, the effective matter in $\Sigma_{y_{0}}$ is restricted to behave either as a radiation-like fluid or as nongravitating matter. Both scenarios are unsatisfactory if we desire to obtain more general cosmologies. The problem is that the embedding $y = y_{0}$ is too rigid.

A more ``flexible" approach, that respects the spatial homogeneity and isotropy of FRW models, is to consider that 4D observers are at rest only  in 3D $(dx^i = 0)$, but moving in 5D. That is we relax the condition $dy = 0$ of section $2.2$ and assume that $y = y(t)$, or  in parametric form 
\begin{equation}
\label{most general embedding}
t = S(\tau), \;\;\;\; y = Y(\tau), 
\end{equation}
where $\tau$ is the proper time. In this approach our spacetime  is recovered on a dynamical 4D hypersurface, which we will denote as  $\Sigma_{Y(\tau)}$,  with local coordinates  $(\tau, r, \theta, \phi)$. In this section we will see that an observer living in $\Sigma_{Y(\tau)}$, who is  unaware of the motion through an empty 5D-dimensional universe, will interpret the expansion or contraction of the universe as if it were governed by an effective matter satisfying some equation of state, which is not necessarily restricted to be $\rho = 3 p$ or $\rho = - 3 p$. 

We consider a foliation of the 5D-dimensional manifold such that (\ref{most general embedding}) is itself a hypersurface of the foliation. Then, for the 5D cosmological line element 
\begin{equation}
\label{General 5D cosmological solution}
d{\cal{S}}^2 = N^2(t, y)dt^2 - A^2(t, y)d\Sigma_{k}^2 + \epsilon \Phi^2(t, y)dy^2,
\end{equation}
the metric induced on every hypersurface of the foliation is (hereafter $\dot{X} \equiv d X/d\tau$; $X_{t} \equiv d X/dt$; $X' \equiv d X/dy)$
\begin{equation} 
ds^2 = \left[N^2(t, y){\dot{S}}^2 + \epsilon \Phi^2(t, y){\dot{Y}}^2\right] d\tau^2 - a^2(\tau)d\Sigma_{k}^2.
\end{equation}
On $\Sigma_{Y(\tau)}$ this metric has to take the  usual FRW form. Therefore we require
\begin{equation}
\label{boundary conditions}
\left[N^2(t, y){\dot{S}}^2 + \epsilon \Phi^2(t, y){\dot{Y}}^2\right]_{_{t = S(\tau), y = Y(\tau)}} = 1, \;\;\;a(\tau) = A(t, y)_{|_{t = S(\tau), y = Y(\tau)}},
\end{equation}
which ensures that $\tau$ is the proper time. It also  shows that 
the functions $S$ and $Y$ are not independent, meaning that the state of movement of $\Sigma_{Y(\tau)}$ is parameterized just by one function of the proper time $\tau$. 
The unit vector ${\hat{n}}^A$ normal to the foliation, and the four-velocity vector $u^{A}$ tangent to the foliation can be written as 
\begin{eqnarray}
\label{unit vectors}
\hat{n}_{A} &=& \frac{s}{\sqrt{1 + \epsilon V^2}}\left( - V N, 0, 0, 0, \Phi\right), \;\;\;\;\hat{n}^{A} = \frac{s}{\sqrt{1 + \epsilon V^2}}\left( - \frac{V}{ N}, 0, 0, 0, \frac{\epsilon}{\Phi}\right),\nonumber \\ 
u_{A} &=& \frac{1}{\sqrt{1 + \epsilon V^2}}\left( N,  0, 0, 0, \epsilon \Phi V\right), \;\;\;\;u^{A} = \frac{1}{\sqrt{1 + \epsilon V^2}}\left(  \frac{1}{ N}, 0, 0, 0, \frac{V}{\Phi}\right).
\end{eqnarray}
Here ${\hat{n}}_{A}{\hat{n}}^{A} = \epsilon$; $u_{A}u^{A} = 1$; $n_{A}u^{A} = 0$; $s = \pm 1$ determines the orientation of the normal\footnote{For $s = 1$ $(s = -1)$ the displacement $n_{A}dx^A$, for fixed $t$, is in the increasing 
 (decreasing) direction of $y$.}, and $V$ is the {\it coordinate} velocity of $\Sigma_{Y(\tau)}$. This follows from the fact that  the displacement ${\hat{n}}_{A}dx^A$ must vanish on every hypersurface of the foliation. Therefore,
\begin{equation}
\label{coordinate velocity}
V = \frac{\Phi dy}{N dt} = \frac{\Phi  \dot{Y}}{N\dot{S}} = \frac{\Phi \dot{Y}}{\sqrt{1 - \epsilon \Phi^2 {\dot{Y}}^2}},
\end{equation}
where  we have used (\ref{boundary conditions}) and assumed $\dot{S} > 0$.

\subsection{Interpretation of   DSV solutions on $\Sigma_{Y(\tau)}$}
We  now apply the above general formulate to elucidate what kind of 4D cosmological models can be obtained as projections of DSV solutions on a moving hypersurface $\Sigma_{Y(\tau)}$.

\subsubsection{Time-dependent DSV solutions}

First, let us show that the effective matter induced on   $\Sigma_{Y(\tau)}$ is no longer pure Weyl radiation as it occurs on $\Sigma_{y_{0}}$,  (\ref{vacuum fluid plus radiation})-(\ref{The function fsubLambda}). In fact,  the extrinsic curvature of the surface $\Sigma_{Y(\tau)}$ is in general non-zero, because  ${\hat{n}}^0 \sim V \neq 0$. In particular, for the  DSV  solutions (\ref{General Davidson el al solution})-(\ref{Phi for the general Davidson solution})    
\begin{equation}
\label{K for dynamical Sigma}
K_{0}^{0} = - s B_{0}A_{t}\left[\frac{\ddot{Y}}{\dot{S}} + \frac{A_{tt}\dot{Y}}{A_{t}}\right], \;\;\;K_{1}^{1} = K_{2}^{2} = K_{3}^{3} = - \frac{s B_{0} A^{2}_{t}\dot{Y}}{A}, \;\;\;
\end{equation}
$K_{04}$ and $K_{44}$ are obtained from $K_{AB}{\hat{n}}^{B} = 0$. Substituting in (\ref{4D Einstein with T and K}) we find that, apart from the Weyl radiation given by $E_{\mu\nu}$,  there are a number of additional terms which do not cancel out, but vanish when $\dot{Y} = 0$ $(V = 0)$. This is a general result valid for any 5D metric with no-dependence of the extra $y$ coordinate. It uncovers the misleading nature of the notion, frequently found in the literature,   that ``radiation is the only kind of matter one can obtain in the induced-matter interpretation as long as the cylinder condition\footnote{Kaluza's cylinder condition sets the  derivatives with respect to the additional coordinate to zero.} is in place."

We now proceed to develop a general expression for the energy density induced on $\Sigma_{Y(\tau)}$. To this end we use (\ref{boundary conditions}), which in the present case  reads

\begin{equation}
\label{Y, S, a for the DSV on moving Sigma}
\left[{\dot{S}}^2 - B_{0}^2 A_{t}^2 {\dot{Y}}^2\right]_{|_{t = S(\tau)}} = 1, \;\;\;a(\tau) = A(S(\tau));
\end{equation}
obtain $S$ as a function of $a(\tau)$ from (\ref{General Davidson el al solution});  calculate ${\dot{S}} = {\dot{a}}(dS/da)$,  and substitute ${\dot{a}}^2 = 8\pi G \rho^{\mbox{eff}} a^2/3 - k$.  The result is
\begin{equation}
\label{condition  for Ydot, original DSV}
B_{0}^2 A_{t}^2 {\dot{Y}}^2 = {\dot{S}}^2 - 1 = \frac{4[8\pi G \rho^{\mbox{eff}} - 3 \epsilon_{\Lambda}\omega^2/4]a^4(\tau) - 3 f_{\Lambda}}{3\delta_{\Lambda}}, \;\;\;\delta_{\Lambda} \geq 0,
\end{equation}
where $\epsilon_{\Lambda}$ and  $f_{\Lambda}$ are the coefficients defined in (\ref{vacuum fluid plus radiation})-(\ref{The function fsubLambda}), and

\begin{equation}
\label{The function delta}
\delta_{\Lambda} = \left\{\begin{array}{cc}
              f_{\Lambda} - 4k a^2 + \omega^2 a^4,    \;\;\;k_{(5)}\Lambda_{(5)} = \frac{3\omega^2}{2},\\
\\
 \; f_{\Lambda}  - 4 k a^2,\;\;\;   \Lambda_{(5)} = 0, \\
\\
  \; f_{\Lambda} - 4 k a^2 - \omega^2 a^4,    \;\;\; k_{(5)}\Lambda_{(5)} = - \frac{3\omega^2}{2}.
               \end{array}
 \right.
\end{equation}
The condition $\delta_{\Lambda} \geq 0$ is necessary to ensure that $S(\tau)$ is a real function. In principle, $f_{\Lambda}$ can have either sign, but models where $a$ can reach  $a = 0$ require $f_{\Lambda} \geq 0$. Finally, from (\ref{Phi for the general Davidson solution}), (\ref{coordinate velocity}) and (\ref{Y, S, a for the DSV on moving Sigma}) we get the desired formula, viz., 

\begin{equation}
\label{equation for rho, DSV}
8\pi G \rho^{\mbox{eff}} = \Lambda_{(4)} + \frac{3 f_{\Lambda}}{4 a^4} + \frac{V^2}{1 - V^2}\left(\frac{3\delta_{\Lambda}}{4 a^4}\right).
\end{equation}
We note that when $V = 0$ we recover (\ref{vacuum fluid plus radiation})-(\ref{The function fsubLambda}). The last term is always positive because $|V| < 1$ $({\dot{Y}}^2 \geq 0)$, and in general is not radiation-like because $\delta_{\Lambda}$ is a function of $a$. 

In principle,  one can find the motion of $\Sigma_{Y(\tau)}$ for any given equation of state. The algorithm is as follows: from the field equations find $a(\tau)$; use (\ref{General Davidson el al solution}) to get $S$ as a function of $a(\tau)$; substitute into (\ref{Y, S, a for the DSV on moving Sigma}) to obtain $V(\tau)$ and $Y(\tau)$; finally, the range of applicability of the model is set by the condition ${\dot{Y}}^2 \geq 0$. 

As an illustration, let us consider the  equation of state $p^{\mbox{eff}} = w \rho^{\mbox{eff}}$ with $w =$ constant, which implies $\rho \sim 1/a^{3(1 + w )}$. Substituting this into (\ref{condition  for Ydot, original DSV}) we find that  ${\dot{Y}}^2 \geq 0$ is satisfied, in the whole range  of $a$,   if  $(\Lambda_{(5)} = 0, w = 1/3)$ or $(\Lambda_{(5)} < 0, w \geq 1/3)$; 
in any other case the applicability is restricted to certain portions of the evolution. For example, in the case of
spatially flat models with $\Lambda_{(4)} = 0$ for which  $a(\tau) \sim \tau^{2/3(w + 1)}$, we find  $S = C \tau^{4/3(w + 1)}$, and\footnote{Please note that $V$ is {\it not} the speed of light $c$. Thus $V = 1$ does not mean that $\Sigma_{Y(\tau)}$ is light-like.} 
\begin{equation}
\label{velocity}
V^2 = 1 - \frac{9(w + 1)^2 \tau^{2(3w - 1)/3(w + 1)}}{16 C^2}.
\end{equation}
The model works well $(0 \leq V^2 \leq 1)$ either in the very early universe $(1/3 < w \leq 1)$ or  at ``late" times $(0 \leq w < 1/3)$.  An interesting by-product of the discussion is that $V = $ constant for $w = 1/3$. Namely,  $V = 0$ for pure Weyl (geometric) radiation, while $V = $ constant $\neq 0$ for a mixture of Weyl radiation, photons and ultra-relativistic matter.

Another approach for constructing 4D cosmological models from (\ref{General Davidson el al solution})-(\ref{Phi for the general Davidson solution}) consists in prescribing the motion of $\Sigma_{Y(\tau)}$. The problem with this is that we have no physical arguments  in support of any particular choice. The only criterion  seems to be ``mathematical simplicity".  Here we just show an interesting  representative example, which arises from the assumption 
\begin{equation}
\label{representative example}
S(\tau) = \tau + C \tau^{4/3(w + 1)},
\end{equation}
 with $0 \leq \tau < \infty$ and $C = $ constant $> 0$. In this case ${\dot{Y}}^2 \geq 0$ $(|V| \leq 1)$ in the whole range of $\tau$, for all $w \in (-1, 1]$. In particular, for $w = 0$,   ${\dot{S}}_{|\tau \rightarrow 0} \rightarrow  1$ in the early universe, which by virtue of (\ref{Y, S, a for the DSV on moving Sigma}) implies  $Y_{|\tau \rightarrow 0} \rightarrow$  constant, i.e., in the early universe $\Sigma_{Y(\tau)} \rightarrow \Sigma_{y_{0}}$. Consequently, at early times the model reduces to a radiation filled universe analogous to the one  considered in (\ref{vacuum fluid plus radiation}), while at late times the effective matter on $\Sigma_{Y(\tau)}$ behaves like {\it cosmological dust}.   
Similar cosmologies  can be generated  on $\Sigma_{Y(\tau)}$ from the  other solutions in (\ref{General Davidson el al solution}).

\subsubsection{Static DSV solutions}

We now consider the case where our spacetime is a dynamical hypersurface  moving in the static background  (\ref{static solution})-(\ref{Transformed General Davidson el al solution}). Here, in addition to the freedom in the choice of $Y(\tau)$ (or $S(\tau)$), we also have a freedom in the signature of the extra dimension. In this case  
the functions $S$ and $Y$ must satisfy
\begin{equation}
\label{relation between S and Y}
B_{0}^2{\cal{A}}'^2 {\dot{S}}^2 + \epsilon {\dot{Y}}^2 = 1, \;\;\;a(\tau) \equiv {\cal{A}}(Y(\tau)),
\end{equation} 
which follow from (\ref{boundary conditions}), and assure that the spacetime metric induced in $\Sigma_{Y(\tau)}$ has the  usual FRW form.  
The matter induced on  $\Sigma_{Y(\tau)}$ is determined by the Einstein equations, namely 
\begin{eqnarray}
\label{induced field equations}
\frac{8\pi G}{3} \rho^{\mbox{eff}} &=& \frac{{\cal{A}}'^2 {\dot{Y}}^2}{{\cal{A}}^2} + \frac{ k}{{\cal{A}}^2}, \nonumber \\
4 \pi G \left(p^{\mbox{eff}} + \frac{\rho^{\mbox{eff}}}{3}\right) &=& - \frac{{\cal{A}}'' {\dot{Y}}^2}{{\cal{A}}} - \frac{{\cal{A}}'\ddot{Y}}{{\cal{A}}}.
\end{eqnarray} 
The last equation shows that $\rho^{\mbox{eff}} = - 3 p^{\mbox{eff}}$,  when $Y =$ constant. However, there is a another solution, namely   ${\cal{A}}(Y(\tau)) \sim \tau$, which generates  Milne's universe (the corresponding $Y$ and $S$ are obtained from (\ref{Transformed General Davidson el al solution}) and (\ref{relation between S and Y}), respectively).  For any other choice the effective matter in $\Sigma_{Y(\tau)}$ behaves as ``regular" gravitating matter. 

Again,  we can derive a general expression for  $\rho^{\mbox{eff}}$ analogous to (\ref{equation for rho, DSV}). Although we omit the formulae, for all three solutions in (\ref{Transformed General Davidson el al solution}) we find that the matter induced on a dynamical hypersurface is compatible with the barotropic equation of state, in the whole range of $\tau$, and  $w \neq 1/3$,  provided  $\Lambda_{(5)} \leq 0$, which is equivalent to   $\epsilon = - 1$ and $\epsilon = 1$ in the hyperbolic and trigonometric solution, respectively. For $w = 1/3$, the extra dimension can  be either spacelike or timelike. As an illustration, let us consider spatially flat models with $\Lambda_{(5)} = 0$. In this case

\begin{equation}
a^2(\tau) = c_{1}Y(\tau) + c_{2} = \left(C_{1} \tau + C_{2}\right)^{4/3(w + 1)},
\end{equation}
where $C_{1}$ and $C_{2}$ are constants of integration. Substituting into (\ref{relation between S and Y}) we get 
\begin{equation}
\label{Sdot is not necessarily positive}
\frac{B_{0}^2 c_{1}^2}{4}{\dot{S}}^2 = a^2 - \frac{16 \epsilon C_{1}^2 a^{3(1 - w)}}{9(1 + w)^2 c_{1}^2}.
\end{equation} 
Since ${\dot{S}}^2 > 0$, and $0 \leq a(\tau) < \infty$, this expression shows that  the extra dimension must be spacelike $(\epsilon = - 1)$ for any $w  \neq 1/3$. For $w = 1/3$, it can be either spacelike or timelike, depending on the choice of the constants.  

Another simple example that demonstrates the diversity of cosmological scenarios, arises from the observation that setting $\dot{S} = 1$ in (\ref{Y, S, a for the DSV on moving Sigma}) requires $Y = Y_{0}$, which gives back the conventional models with radiation-like effective matter (\ref{vacuum fluid plus radiation}).  In static DSV solutions the assumption   $\dot{S} = 1$ is inconsistent with  $Y = Y_{0}$, but generates  an ample variety of collapsing  and bouncing  cosmological models. In fact,  the scale factor  resulting from the integration of (\ref{relation between S and Y})  can be written in the parametric form

\begin{equation}
\label{Sdot = 1, epsilon = - 1}
a = \frac{B_{0}c_{1}}{2} \sin{\eta}, \;\;\; \tau - \tau_{0} = \frac{B_{0}^2 c_{1}}{4}\left(\eta - \frac{1}{2}\sin{2 \eta}\right), \;\;\;\mbox{for} \;\;\;\epsilon = - 1,
 \end{equation}
where $\tau_{0}$ is a constant of integration, which can be set equal to zero. Thus, despite the fact that the universe is spatially flat, for $\epsilon = -1$ it  recollapses in a finite proper time: $a(\tau)$ grows from zero at $\tau = 0$ $(\eta = 0)$ to a maximum value $a = {B_{0}c_{1}}/{2}$, which is reached when  $\tau =  B_{0}^2 c_{1}\pi/8$ $(\eta = \pi/2)$, and decreases again to zero at $\tau =  B_{0}^2 c_{1}\pi/4$.  For small values of $a$, we assume $\eta \ll 1$. Then $a \sim \eta$, $\tau \sim \eta^3$, so that  $a \sim \tau^{1/3}$, which corresponds to a fluid with a stiff equation of state, i.e., $p^{\mbox{eff}} = \rho^{\mbox{eff}}$.

In a similar way, for  $\epsilon = 1$ the solution can be written as 
\begin{eqnarray}
\label{Sdot = 1, epsilon = 1}
a =  \frac{B_{0}c_{1}}{2}\cosh{\eta}, \;\;\; \tau - \tau_{0} = \frac{B_{0}^2 c_{1}}{4}\left(\eta + \frac{1}{2}\sinh{2 \eta}\right), \;\;\;\mbox{for}\;\;\;\epsilon = 1. 
\end{eqnarray}
Here the scale factor  changes monotonically, increasing from $a = B_{0} c_{1}/2$, at $\tau = 0$ $(\eta = 0)$,  to infinity for $\tau \rightarrow \infty$ $(\eta \rightarrow \infty)$. For large values of $\tau$ we find $a \sim \tau^{1/2}$, which corresponds to radiation, $\rho^{\mbox{eff}} = 3p^{\mbox{eff}}$.

To sum up, in this section we have shown that an observer riding in a hypersurface $\Sigma_{Y(\tau)}$ can perceive a rich variety of cosmological scenarios, including cosmological models where the induced matter satisfies the barotropic equation of state, not only radiation-like or nongravitating matter.

\section{Our universe as a moving brane in a static DSV universe}

In the last section, within the context of IM, we have obtained  cosmological models in 4D  as projections of the $5$-dimensional DSV  solutions on a moving hypersurface $\Sigma_{Y(\tau)}$.
The aim of this section is to extend the discussion to embrace the so-called braneworld cosmological models. In these models gravity propagates in all $5$-dimensions, whereas particles and fields are confined to a singular 4D hypersurface (the brane), thereby accounting for their  relatively greater strength. 

The metric $g_{AB}$ induced on the hypersurfaces of the foliation, which is defined by the orthonormal vectors (\ref{unit vectors}), is given by
\begin{equation}
\label{metric on the hypersurfaces of the foliation}
g_{AB} = \gamma_{AB} - \epsilon \hat{n}_{A}\hat{n}_{B}, \;\;\;g_{AB}\hat{n}^{B} = 0.
\end{equation} 
Thus, using (\ref{General 5D cosmological solution}) and (\ref{unit vectors}) we obtain
\begin{eqnarray}
g_{00} &=& \frac{N^2}{1 + \epsilon V^2}\;\;\;\;g_{ij} = \gamma_{ij}, \nonumber \\
g_{04} &=& \epsilon\left(\frac{V}{N}\right)g_{00}, \;\;\;\;g_{44} = \left(\frac{V}{N}\right)^2g_{00},
\end{eqnarray}
where,  from (\ref{static solution})
\begin{equation}
\label{N}
N \equiv  B_{0} \left(\frac{d{\cal{A}}}{dy}\right).
\end{equation}
The extrinsic curvature is 
\begin{eqnarray}
\label{extrinsic curvature for the moving brane}
K_{00} &=& s \; \frac{\epsilon N^2}{\left(1 + \epsilon V^2\right)^{3/2}}\left[\frac{N'}{N} - \frac{\epsilon V V'}{{1 + \epsilon V^2}}\right], \nonumber \\
K_{1}^{1} &=& K_{2}^{2} = K_{3}^{3} = s\; \frac{\epsilon {\cal{A}}'}{{\cal{A}}\sqrt{1 + \epsilon V^2}}, \nonumber \\
K_{04} &=& \epsilon \left(\frac{V}{N}\right)K_{00}, \;\;\;          K_{44} = \left(\frac{V}{N}\right)^2 K_{00},
\end{eqnarray}
with trace 
\begin{equation}
\label{trace of K}
K =  \frac{s \epsilon}{\sqrt{1 + \epsilon V^2}}\left[\frac{N'}{N} + \frac{3{\cal{A}}'}{{\cal{A}}} - \frac{ \epsilon V V'}{1 + \epsilon V^2}\right].
\end{equation}
In the present case the coordinate velocity (\ref{coordinate velocity}) reduces to $V = \dot{Y}/\sqrt{1 - \epsilon {\dot{Y}}^2}$. 
Therefore, in the above equations   
\begin{equation}
\label{V and V' in terms of Ydot}
\sqrt{1 + \epsilon V^2} = \frac{1}{\sqrt{1 - \epsilon {\dot{Y}}^2}}, \;\;\; V V' = \frac{\ddot{Y}}{(1 - \epsilon {\dot{Y}}^2)^2}, 
\end{equation}
where we have used that $V' = \dot{V}/\dot{Y}$ \cite{Campos Maartens}.

\subsection{Matter on the brane}

The metric $g_{AB}$ is continuous across the brane but there is a jump in the extrinsic curvature $({K_{AB}}_{|s = 1} = - {K_{AB}}_{|s = - 1})$. Israel's boundary conditions \cite{Israel} relate this jump to the brane energy-momentum tensor, viz.,   
\begin{equation}
\label{EMT on the brane}
T_{AB}  + \sigma g_{AB} = - \frac{2 \epsilon}{k_{(5)}^2}\left(K_{AB} - g_{AB}K\right),
\end{equation}
where $K_{AB} = {K_{AB}}_{|s = + 1}$; $\sigma$ is the tension of the brane, which is interpreted as the vacuum energy density, and $T_{AB}$ represents the energy-momentum tensor of ordinary matter in the brane. For perfect fluid it is
\begin{equation}
\label{perfect fluid on the brane}
T_{AB} = (\rho + p)u_{A}u_{B} - p g_{AB},
\end{equation} 
where $\rho$ and $p$ are the energy density and isotropic pressure measured by an observer with velocity $u^{A}$.

If we substitute (\ref{Transformed General Davidson el al solution}), (\ref{N}), (\ref{V and V' in terms of Ydot}) into (\ref{EMT on the brane})-(\ref{perfect fluid on the brane}),  and set $y = Y(\tau)$, then we obtain the braneworld matter  in terms of the embedding function $Y(\tau)$,  and its first and second derivatives.  However, in cosmological models, what we need is to find explicit expressions relating the matter in 4D to the dynamics of the scale factor $a(\tau)$. To accomplish this goal we have to specify  $Y$ as a function of  $a(\tau)$.  
\begin{enumerate}
\item From the first solution in (\ref{Transformed General Davidson el al solution}) we obtain
\begin{equation}
\label{explicit expression for cosh}
\cosh{\omega Y} = \frac{c_{1}[\omega^2 a^2(\tau) - 2k\epsilon] - c_{2} \Delta_{h}}{\omega^2 (c_{1}^2 - c_{2}^2)},
\end{equation}
with $\Delta_{h} = \sqrt{   [\omega^2 a^2(\tau) - 2k\epsilon]^2 - \omega^4(c_{1}^2 - c_{2}^2)}$.  The expression for $\sinh{\omega Y}$ is obtained by changing $c_{1} \leftrightarrow c_{2}$ in (\ref{explicit expression for cosh}), but keeping $\Delta_{h}$. We note that $\cosh{\omega Y}$ and $\sinh{\omega Y}$  remain finite for $c_{1} = c_{2}$. 
\item From the second solution in (\ref{Transformed General Davidson el al solution}) we get 
\begin{eqnarray}
\label{Y for the second solution}
Y(\tau) &=& \frac{\epsilon}{2 k}\left\{c_{1} - \sqrt{c_{1}^2 + 4\epsilon k[c_{2} - a^2(\tau)]}\right\}, \;\;\;k \neq 0.
\end{eqnarray}
For $k = 0$, $Y(\tau) = [a^2(\tau) - c_{2}]/c_{1}$.
\item Finally, we note that the third solution in (\ref{Transformed General Davidson el al solution}) is formally obtained from the first one under the substitution  $\omega \rightarrow i \omega$, $c_{2} \rightarrow i c_{2}$. Therefore we have 

\begin{equation}
\label{explicit expression for cos}
\cos{\omega Y} = \frac{c_{1}[\omega^2 a^2(\tau) + 2k\epsilon] - c_{2}\Delta_{t}  }{\omega^2 (c_{1}^2 + c_{2}^2)},
\end{equation}
where $\Delta_{t} = \sqrt{ \omega^4(c_{1}^2 + c_{2}^2) - [\omega^2 a^2(\tau) + 2k\epsilon]^2 }$.  The expression for $\sin{\omega Y}$ is obtained from this equation by changing $c_{1} \rightarrow c_{2}$, $c_{2} \rightarrow - c_{1}$. 
\end{enumerate}
Substituting (\ref{explicit expression for cosh})-(\ref{explicit expression for cos}) into the original equations for braneworld matter in terms of  $Y(\tau)$, we obtain (to simplify the presentation we omit cumbersome  intermediate calculations)
\begin{equation}
\label{Generalized Friedmann equation}
 3\left(\frac{\dot{a}}{a}\right)^2 = \left\{\begin{array}{cc}
              \frac{3 \epsilon \omega^2}{4}  - \frac{\epsilon k_{(5)}^4}{12}(\rho + \sigma)^2 - \frac{3k}{a^2} + \frac{3 \epsilon[4k^2 -(c_{1}^2 - c_{2}^2)\omega^4]}{4\omega^2 a^4},    \;\;\;\mbox{for}\;\;\;k_{(5)}^2\Lambda_{(5)} = \frac{3\epsilon \omega^2}{2},\\
\\
 \;\; - \frac{\epsilon k_{(5)}^4}{12}(\rho + \sigma)^2 - \frac{3k}{a^2} + \frac{3\epsilon}{a^4}\left(\frac{c_{1}^2}{4} + \epsilon k c_{2}\right),\;\;\;   \mbox{for}\;\;\;k_{(5)}^2\Lambda_{(5)} = 0, \\
\\
   - \frac{3 \epsilon \omega^2}{4}  - \frac{\epsilon k_{(5)}^4}{12}(\rho + \sigma)^2 - \frac{3k}{a^2} -  \frac{3 \epsilon[4k^2 -(c_{1}^2 + c_{2}^2)\omega^4]}{4\omega^2 a^4}, \;\;\; \mbox{for} \;\;\; k_{(5)}^2\Lambda_{(5)} = - \frac{3 \epsilon \omega^2 }{2}.
               \end{array}
 \right.
\end{equation}
By means of direct calculation we have checked  that the three models satisfy the conservation equation
\begin{equation}
\label{conservation equation}
\left(\dot{\rho} + \dot{\sigma}\right) + 3 \frac{{\dot{a}}}{a}\left(\rho + p\right) = 0,
\end{equation}
so we need not to have an  explicit expression for $(p - \sigma)$. However, it may be worth emphasizing that this conservation equation holds regardless of the choice of functional dependence of $\sigma(\tau)$, although it is usually taken as $\sigma = \sigma_{0} = $ constant.   

\subsubsection{The generalized Friedmann equation}

Following a notation that now is standard, the above solutions can be written as

\begin{equation}
\label{generalized Friedmann equation from the above  solutions}
3 \left(\frac{{\dot{a}}}{a}\right)^2 = \Lambda_{(4)} + 8\pi G \rho - \frac{3 k}{a^2} - \frac{\epsilon k_{(5)}^4}{12}\rho^2  + \frac{{\cal{C}}}{a^4}
\end{equation}
where
\begin{equation}
\label{definition of G and Lambda4}
8\pi G = - \frac{\epsilon k_{(5)}^4 \sigma}{6},\;\;\;\;\;\Lambda_{(4)} = \left\{\begin{array}{cc}
              \frac{3 \epsilon}{4}\left[\omega^2 - \frac{1}{9}k_{(5)}^2 \sigma^2\right],    \;\;\;\mbox{for}\;\;\;k_{(5)}^2\Lambda_{(5)} = \frac{3\epsilon \omega^2}{2},\\
\\
 \;\;\;\;\;\;\;\;\;\; - \epsilon \frac{k_{(5)}^4 \sigma^2}{12},\;\;\;   \mbox{for}\;\;\;k_{(5)}^2\Lambda_{(5)} = 0, \\
\\
   - \frac{3 \epsilon}{4}\left[\omega^2 + \frac{1}{9}k_{(5)}^2 \sigma^2\right], \;\;\; \mbox{for} \;\;\; k_{(5)}^2\Lambda_{(5)} = - \frac{3 \epsilon \omega^2 }{2}.
               \end{array}
 \right.
\end{equation}
and

\begin{equation}
\label{calC for the Generalized Friedmann equation}
 {\cal{C}} = \left\{\begin{array}{cc}
              \frac{3 \epsilon \left[4 k^2 - \left(c_{1}^2 - c_{2}^2\right)\omega^4\right]}{4 \omega^2},    \;\;\;\mbox{for}\;\;\;k_{(5)}^2\Lambda_{(5)} = \frac{3\epsilon \omega^2}{2},\\
\\
 \;\; \frac{3 \epsilon (c_{1}^2 + 4 \epsilon k c_{2})}{4},\;\;\;   \mbox{for}\;\;\;k_{(5)}^2\Lambda_{(5)} = 0, \\
\\
   - \frac{3 \epsilon \left[4 k^2 - \left(c_{1}^2 + c_{2}^2\right)\omega^4\right]}{4 \omega^2}, \;\;\; \mbox{for} \;\;\; k_{(5)}^2\Lambda_{(5)} = - \frac{3 \epsilon \omega^2 }{2}.
               \end{array}
 \right.
\end{equation}
The first three terms in the  r.h.s. of (\ref{generalized Friedmann equation from the above  solutions}) give the standard cosmology, while the fourth and fifth terms  are local and non-local higher dimensional modifications to general relativity, respectively. In particular, the term ${\cal{C}}/a^4$ can be interpreted as an effective (Weyl) radiation coming from the bulk to the brane.
The first solution in (\ref{generalized Friedmann equation from the above  solutions})-(\ref{definition of G and Lambda4}) with $\epsilon = - 1$ is in agreement with the well-known generalized Friedmann equation \cite{Binetruy}-\cite{Ida}, which has widely been used and discussed in the braneworld literature.  In this equation one can always fine-tune the parameters $\omega$ and $\sigma$ is such a way as to set the 4D effective cosmological constant $\Lambda_{(4)} $ equal to zero. 

However, this is not so for the second and third solutions in (\ref{generalized Friedmann equation from the above  solutions})-(\ref{definition of G and Lambda4}). For the solution with $\Lambda_{(5)} = 0$ this is not surprising. But,  in the third solution $\Lambda_{(4)}$ cannot be set equal to zero either,  despite of the fact that $\Lambda_{(5)} \neq 0$. 

\subsection{Consistency relations for the embedding}

First, we have to verify the positivity of ${\dot{S}}^2$. In fact, from (\ref{Sdot is not necessarily positive}) we have learned that ${\dot{S}}^2$ is not automatically positive; in principle the requirement ${\dot{S}}^2 > 0$ may lead to some specific physical restrictions. 
In the present case one can show that ${\dot{S}}^2 > 0$, for all three solutions, without imposing restrictions whatsoever. 

Indeed, from (\ref{relation between S and Y}) it follows that $B_{0}^2 {{\cal{A}}'}^2 {\dot{S}}^2 = (1 - \epsilon {\dot{Y}}^2)$. Now we use (\ref{explicit expression for cosh}), (\ref{Y for the second solution}), (\ref{explicit expression for cos}) to obtain $\dot{Y}$ in terms of $a$ and $\dot{a}$. Next, we eliminate $\dot{a}$ in $\dot{Y}$ by utilizing (\ref{Generalized Friedmann equation}). After a long but straightforward calculation we get
\begin{equation}
B_{0}^2 {{\cal{A}}'}^4 {\dot{S}}^2 = \frac{k_{(5)}^2}{36}\left(\rho + \sigma\right)^2 a^2(\tau),
\end{equation}
for the three cases, regardless of the choice of $\epsilon$, $k$ and ${\cal{C}}$.

Second, we have to make sure that $Y(\tau)$ is a real function for all values of $\tau$. This requires the quantities under the roots in (\ref{explicit expression for cosh}), (\ref{Y for the second solution}) and (\ref{explicit expression for cos}) to be non-negative. Using (\ref{calC for the Generalized Friedmann equation}), this condition can be expressed as 

\begin{eqnarray}
\label{geometrical constraints}
\omega^2 a^4 - 4 k \epsilon a^2 + \frac{4 \epsilon {\cal{C}}}{3} &\geq 0&, \;\;\;\mbox{for}\;\;\;k_{(5)}^2\Lambda_{(5)} = \frac{3\epsilon \omega^2}{2},\nonumber \\
\epsilon \left[{\cal{C}} - 3k a^2\right] &\geq 0&, \;\;\;\mbox{for}\;\;\;k_{(5)}^2\Lambda_{(5)} = 0,\nonumber \\
\omega^2 a^4 + 4 k \epsilon a^2 - \frac{4 \epsilon {\cal{C}}}{3} &\leq 0&, \;\;\;\mbox{for}\;\;\;k_{(5)}^2\Lambda_{(5)} = - \frac{3\epsilon \omega^2}{2}.
\end{eqnarray}
These equations, impose geometrical  constraints on the evolution of the scale factor $a$, for every given set  $(\epsilon, k, {\cal{C}})$. They indicate what kind of braneworld cosmological models are, in principle,  compatible with the  embeddings under consideration. 
However, physical requirements, as the energy conditions on $\rho$ and $p$ and  ${\dot{a}}^2 \geq 0$, demand more stringent constraints on the evolution of the scale factor.

\subsubsection{Possible cosmological scenarios}

We now proceed to  examine, in some detail,  the 4D cosmological models allowed by (\ref{geometrical constraints}) when  the extra dimension is spacelike $(\epsilon = - 1)$.

The models with $k_{(5)}^2\Lambda_{(5)} = - {3\omega^2}/{2}$ are  specially interesting because, as we mentioned above, they correspond  to the generalized Friedmann equation discussed in braneworld cosmologies \cite{Binetruy}-\cite{Ida}.  
 In this case the inequality $\omega^2 a^4 + 4 k  a^2 - {4  {\cal{C}}}/{3} \geq 0$ has several solutions corresponding to the various values of ${\cal{C}}$ and $k$: 

\begin{enumerate}
\item {\underline {$k = - 1$}}: For $- \infty < {\cal{C}} \leq - 3/\omega^2$, there are no restrictions on $a$, i.e., the above inequality is satisfied for {\it all} values of $a$.  For $-3/\omega^2 <  {\cal{C}} < 0$, the allowed range of $a$ is (assuming $a$ positive) either 

\begin{equation}
0 \leq a(\tau) \leq \frac{\sqrt{2}}{|\omega|}\left(1 - \sqrt{1 + \omega^2 {\cal{C}}/3}\right)^{1/2}, \;\;\;\mbox{or}\;\;\;a(\tau) \geq \frac{\sqrt{2}}{|\omega|}\left(1 + \sqrt{1 + \omega^2 {\cal{C}}/3}\right)^{1/2}.
\end{equation}
For ${\cal{C}} \geq 0$, $a$ is restricted to the  region defined by the second inequality.

\item {\underline { $k = (0, 1)$}}:  For ${\cal{C}} \leq 0$, there are no restrictions on $a$. For ${\cal{C}} > 0$ we obtain
\begin{equation}
\label{possible cosm. scenarios k = 0, 1}
a(\tau) \geq \left(4{\cal{C}}/3\omega^2\right)^{1/4}, \;\;\;\mbox{and}\;\;\;a(\tau) \geq \frac{\sqrt{2}}{|\omega|}\left(\sqrt{1 + \omega^2 {\cal{C}}/3} - 1\right)^{1/2},
\end{equation}
for $k = 0$ and $k =  1$, respectively. We note that $a(\tau)$ can never reach zero, i.e. there is no big-bang, unless ${\cal{C}} = 0$.

\end{enumerate}

Summing-up, the case where $k_{(4)}^2\Lambda_{(5)} = - 3 \omega^2/2$ allows recollapsing, bouncing and ever-expanding cosmological models. It is not difficult to see that similar scenarios are possible when $\Lambda_{(5)} = 0$. However, for $k_{(4)}^2\Lambda_{(5)} =  3 \omega^2/2$ {\it {all the models are recollapsing}}.

\medskip

Let us now examine the physics in the bulk. A simple inspection of (\ref{static solution})-(\ref{Transformed General Davidson el al solution}) shows that $g_{tt}(y) = \infty$ at  ${\cal{A}}(y) = 0$. To analyze this in more detail we calculate the Kretschmann scalar $I = R_{ABCD}R^{ABCD}$. We obtain  

\begin{equation}
\label{Kretschmann scalar}
I = 40 \alpha^2 + \frac{72 \beta^2}{{\cal{A}}^8},
\end{equation} 
where
\begin{equation}
\label{definition of alpha and beta}
\alpha = \left(\frac{\omega^2}{4},\;  0,\;  - \frac{\omega^2}{4}\right), \;\;\;\beta = \frac{4{\cal{C}}}{3} = \left[\frac{\omega^4 (c_{1}^2 - c_{2}^2) - 4 k^2}{4 \omega^2},\;\;  \frac{ 4 k c_{2} - c_{1}^2}{4},\;\;  - \frac{\omega^4 (c_{1}^2 + c_{2}^2) - 4 k^2}{4 \omega^2}\right],
\end{equation}
for $k_{(5)}^2 \Lambda_{(5)} = - 3\omega^2/2$;  $k_{(5)}^2 \Lambda_{(5)} = 0$ and $k_{(5)}^2 \Lambda_{(5)} = 3\omega^2/2$, respectively. The above suggests the introduction of 
the dimensionless coordinate $z$, viz., 
\begin{equation}
\label{coordinate transformation to the Sitter}
z = {\cal{A}}.
\end{equation}
In terms of $z$ the static DSV solutions (\ref{static solution})-(\ref{Transformed General Davidson el al solution}), with $\epsilon = - 1$, become
\begin{equation}
\label{de Sitter form for DVS}
dS^2 = B_{0}^2\left(k + \alpha z^2 - \frac{\beta}{z^2}\right)dt^2 - \frac{d z^2}{\left(k + \alpha z^2 - \beta/z^2\right)} - z^2 d \Sigma_{k}^2,
\end{equation}
which is the five-dimensional analogue of the Schwarzschild-de Sitter spacetime.   
It shows that for  $\beta = 0$ $({\cal{C}} = 0)$ the bulk is singularity-free; it is either dS$_{5}$ or   AdS$_{5}$, depending on whether $\alpha < 0$ or $\alpha > 0$. However, for $\beta \neq 0$ the bulk is singular at $z = 0$. For several values of the constants, the singularity is covered by a horizon located at  $z = z_{h}$.

For $\alpha > 0$  

\begin{equation}
\label{horizons for the first solution}
z_{h}^2 = \left\{\begin{array}{cc}
              \sqrt{c_{1}^2 - c_{2}^2} - 2/\omega^2,    \;\;\;k = 1,\\
\\
 \; \sqrt{c_{1}^2 - c_{2}^2} ,\;\;\;   k  = 0, \\
\\
  \; \sqrt{c_{1}^2 - c_{2}^2} + 2/\omega^2,    \;\;\; k = -1.
               \end{array}
 \right.
\end{equation}
Alternative formulas for $z_{h}^2$ can be obtained by using (\ref{definition of alpha and beta}) for expressing $\sqrt{c_{1}^2 - c_{2}^2}$ in terms of ${\cal{C}}$. We find that $z_{h}^{2} > 0$ requires ${\cal{C}} > 0$ for $k = (0, 1)$ and ${\cal{C}} > - (3/4 \omega^2)$ for $k = - 1$. For these parameters the scale factor $a(\tau)$ can never vanish because it is bounded from bellow. 
A similar situation occurs for $\alpha = 0$ and  $k = 1$. 

Our analysis shows that ever-expanding big-bang cosmological models on the effective  $4D$ brane   require $\alpha \geq 0 $ $(\Lambda_{(5)} \leq 0)$ and ${\cal{C}} = 0$ $(\beta = 0)$, which in turn, by virtue of (\ref{Kretschmann scalar}),  demand  the bulk to be  free of singularities (not a black hole in $5D$). In this context, the singularity in $4D$ can be interpreted as a consequence of the topological separation of our universe from the $5D$ bulk \cite{topological separation}.

\section{Summary}

We have studied the classical Davidson-Sonnenschtein-Vozmediano cosmological solutions, originally obtained and interpreted in the context of 5-dimensional Kaluza-Klein theory with cilindricity, where the effective metric in 4D is constructed by a factorization technique (\ref{4D metric in Kaluza-Klein with cilindricity}).    Nowadays, such a factorization is not required and the extra dimension is not assumed to be compactified. The two versions of 5D relativity in vogue, namely induced matter theory and membrane theory, employ a 5D Kaluza-Klein type of metric but identify our spacetime with some  4D hypersurface embedded in  5D. 

In section $2$, by employing Campbell's theorem, which is the fundamental mathematical support of induced-matter theories,  we analyzed the physics induced on  a 4D hypersurface $y = y_{0}$ = constant.  Also, exploiting the symmetry of the 5D metrics, we  constructed the static counterpart to Davidson-Sonnenschtein-Vozmediano  solutions. The equations of state of the matter induced in 4D is $\rho = 3 p$ in the FRW case, and $\rho = - 3p$ in the static case, corresponding to radiation-like and nongravitating matter, respectively.  

In section $3$ we considered the most general embedding compatible with spatial homogeneity and isotropy. The spacetime was identified with a dynamical  hypersurface $\Sigma_{Y(\tau)}$ defined by one function of the proper time. We found that, an observer living in $\Sigma_{Y(\tau)}$, who is  unaware of her/his motion through an empty 5-dimensional universe, will interpret the scale factor $a(\tau)$ as if it were governed by an effective matter satisfying some equation of state, which is not necessarily restricted to be radiation-like or nongravitating. In fact, we discussed  a number of cosmological scenarios, which include ever-expanding, collapsing and bouncing models,  with different equations of state including the barotropic one. 

In section $4$ we used the braneworld paradigm to embed our spacetime as a dynamical 4D hypersurface in a static DSV universe. We considered  three possible cases, viz.,   
$k_{(4)}^2\Lambda_{(5)} = 3\epsilon \omega^2/2$, $\Lambda_{(5)} = 0$, $k_{(4)}^2\Lambda_{(5)} = - 3\epsilon \omega^2/2$, which require different consistency relations for the embedding. As a consequence, although  the generalized Friedmann equation  in 4D, looks the same in all cases, they represent distinct physical scenarios. The most notorious difference is between the cases with $k_{(4)}^2\Lambda_{(5)} = 3\epsilon \omega^2/2$
and $k_{(4)}^2\Lambda_{(5)} = - 3\epsilon \omega^2/2$. The first case, which for $\epsilon = -1$ gives back previous results in the literature \cite{Binetruy}-\cite{Ida}, is compatible  with a wide range of cosmological and  $\Lambda_{(4)}$ can be set equal to zero. The second case, however, is compatible only with recollapsing models and $\Lambda_{(4)}$ cannot be set equal to zero. 

In conclusion, here we have obtained a number of 4D cosmological models  as projections of DSV cosmological solutions on some 4-dimensional hypersurface.  This is the first time in the literature that these solutions are used in a systematic way to provide a different formulation of 4D cosmologies. Therefore, our approach is complementary to the usual formalism employed  in  IM \cite{general} and BW \cite{Binetruy}-\cite{Ida}, and shows that Davidson-Sonnenschtein-Vozmediano  solutions are compatible with the notion that our universe can be an evolving  4D hypersurface embedded in a 5-dimensional world.

\end{document}